\documentclass[11pt,twocolumn,tight,times]{aastex631}
\usepackage{amsmath, mathrsfs}

\newcommand{\appendblock}[1]{#1}

\newcommand{\msun}{{M_\odot}}
\newcommand{\pgal}{{\rm \,gal^{-1}}}
\newcommand{\pyr}{{\rm \,yr^{-1}}}
\newcommand{\gyr}{{\rm\,Gyr}}
\newcommand{\kms}{{\rm \,km\,s^{-1}}}
\newcommand{\fudge}{{f_\omega}}
\newcommand{\consp}{_{\rm consp}}
\newcommand{\peak}{_{\rm peak}}
\newcommand{\calE}{\mathcal{E}}
\newcommand{\lnR}{\ln R_0^{-1}}
\newcommand{\lrT}{_{{\rm lr},T}}
\newcommand{\swl}{_{\rm swl}}
\newcommand{\bh}{_{\rm BH}}
\newcommand{\lc}{_{\rm lc}}
\newcommand{\lr}{_{\rm lr}}
\newcommand{\rmb}{_{\rm b}}
\newcommand{\rmc}{_{\rm c}}
\newcommand{\rme}{_{\rm e}}
\newcommand{\rmh}{_{\rm h}}
\newcommand{\rmr}{_{\rm r}}
\newcommand{\rmt}{_{\rm t}}
\newcommand{\rmeq}{_{\rm eq}}

\newcommand{\Flc}{{F^{\rm lc}}}
\newcommand{\Fdrain}{{F^{\rm drain}}}
\newcommand{\scrFlc}{{\mathscr{F}^{\rm lc}}}
\newcommand{\scrFdrain}{{\mathscr{F}^{\rm drain}}}
\newcommand{\scrFlcPL}{{\mathscr{F}^{\rm lc,PL}}}
\newcommand{\scrFdrainPL}{{\mathscr{F}^{\rm drain,PL}}}
\newcommand{\mbhsigmarelax}{{\sigma\rmh^3 M\bh^{-1}}}
\newcommand{\mbhsigmarelaxunit}{{\sigma_{\rm h,200}^3 M_{\rm BH,8}^{-1}}}
\newcommand{\mbhsigmadrain}{{\kappa(M\bh,\sigma\rmh)}}
\newcommand{\Rlrterm}{{(R\lrT/0.1)^{\frac{2}{5}\alpha-\frac{1}{5}}}}

\shorttitle{Correlations of TDE rates with MBH and Host Galaxy Properties}
\shortauthors{Chen, Yu, \& Lu}

\begin{document}

\title{The correlations of stellar tidal disruption rates with properties of
massive black holes and their host galaxies}

\author[0000-0001-5393-9853]{Yunfeng Chen} 
\affiliation{School of Astronomy and Space Science, University of Chinese
Academy of Sciences, Beijing 100049, China}
\affiliation{National Astronomical Observatories, Chinese Academy of Sciences,
Beijing, 100101, China; luyj@nao.cas.cn}

\affiliation{Kavli Institute for Astronomy and Astrophysics, and School of
Physics, Peking University, Beijing, 100871, China; yuqj@pku.edu.cn}

\author[0000-0002-1745-8064]{Qingjuan Yu}
\affiliation{Kavli Institute for Astronomy and Astrophysics, and School of
Physics, Peking University, Beijing, 100871, China; yuqj@pku.edu.cn}


\author[0000-0002-1310-4664]{Youjun Lu} 
\affiliation{National Astronomical Observatories, Chinese Academy of Sciences,
Beijing, 100101, China; luyj@nao.cas.cn}

%
\affiliation{School of Astronomy and Space Science, University of Chinese
Academy of Sciences, Beijing 100049, China}

\correspondingauthor{Qingjuan Yu}

\begin{abstract}
Stars can be either disrupted as tidal disruption events (TDEs) or swallowed as
a whole by massive black holes (MBHs) at galactic centers when they approach
sufficiently close to these MBHs. In this work, we investigate the correlations of such
stellar consumption rates with both the MBH mass $M\bh$ and the inner slope of
the host galaxy mass density distribution $\alpha$.  We introduce a simplified
analytical power-law model with a power-law stellar mass density distribution
surrounding MBHs and separate the contributions of two-body relaxation and
stellar orbital precession for the stellar orbital angular momentum evolution
in nonspherical galaxy potentials.  The stellar consumption rates derived from
this simplified model can be well consistent with the numerical results
obtained with a more realistic treatment of stellar distributions and dynamics
around MBHs, providing an efficient way to estimate TDE rates. The origin of
the correlations of stellar consumption rates with $M\bh$ and $\alpha$ are
explained by the dependence of this analytical model on those MBH/host galaxy
properties and by the separation of the stellar angular momentum evolution
mechanisms.  We propose that the strong positive correlation between the rates
of stellar consumption due to two-body relaxation and $\alpha$ provides one
interpretation for the overrepresentation of TDEs found in some rare
E+A/poststarburst galaxies. We find high TDE rates for giant stars, up 
to those for solar-type stars. The understanding of the origin of the
correlations of the stellar consumption rates will be necessary for obtaining
the demographics of MBHs and their host galaxies via TDEs.
\end{abstract}


\section{Introduction}
\label{sec:introduction}

Tidal disruption event (TDE) happens when a star travels too close to a massive
black hole (MBH) so that the tidal force exerted on the star by the MBH
surpasses the star's self-gravity, causing the star to be ripped apart and
disrupted, accompanied by luminous flares due to subsequent accretion of the
stripped stellar material \citep[e.g.,][]{Hills75, Rees88}. TDEs may frequently
occur at the centers of galaxies, as observations have shown the ubiquitous
existence of MBHs (with mass ${\sim}10^6$--$10^{10}\msun$) at galactic nuclei
(e.g., \citealt{Ferrarese00, Gebhardt00, Tremaine02, KH13, Graham16}). These
TDEs provide rich electromagnetic signals, which help to study the relativistic
effects, accretion physics, formation of radio jets and interior structure of
torn stars, etc. \citep{Komossa15, Alexander17, Gezari21}. 

TDEs illuminate those dormant MBHs, and the event rates in individual galaxies
depend on the MBH properties (e.g., \citealt{MT99}, hereafter MT99;
\citealt{Wang04, Ivanov05, Kesden12, Fialkov17}). Therefore, they can serve as
powerful probes of the MBH demographics, such as the MBH mass, spin, binarity,
and the occupation fraction of MBHs in galaxies \citep{Gezari21}. Depending on
the event rate, the consumption of intruding stars may provide an important
channel for the growth of MBHs in the centers of galaxies \citep{Freitag02,
Yu03, Brockamp11, Alexander17} and even has significant effects on the MBH spin
evolution \citep{Zhang19}, especially for those relatively less massive ones.

TDEs are first discovered as energetic transients in the soft X-ray band by
archival searches of the ROSAT All-Sky Survey data \citep{Bade96, Grupe99,
Komossa99, Greiner00} and later found via searches by multi-wavelength (e.g.,
UV, optical, X-ray, and gamma-ray) observations \citep{Alexander17, Gezari21}.
Over the past decade, statistically significant samples of TDEs have begun to
accumulate, largely owing to the rapid advancement of the optical time domain
surveys, such as PTF \citep{Law09, Rau09}, ASAS-SN \citep{Shappee14}, Pan-STARRS
\citep{Chambers16}, and ZTF \citep{Bellm19}. For example, a recent census by
\citet{Gezari21} listed $56$ TDE candidates, among which around two thirds are
discovered by the wide-field optical time domain surveys. Based on different
samples of TDE candidates, the TDE event rate is estimated to be
${\sim}10^{-5}$--$10^{-4}\pyr\pgal$ (e.g., \citealt{Donley02, Esquej08,
Maksym10, WZKetal12, vanVelzen14, Khabibullin14, Auchettl18, vanVelzen18}).

Some recent studies on the TDE host galaxies reveal intriguingly that TDEs
prefer those rare E+A/poststarburst galaxies \citep{Arcavi14, French16,
French17, LawSmith17, Graur18, Hammerstein21}. After removing the accountable
selection effects (e.g., MBH mass, redshift completeness, strong active galactic
nucleus presence, bulge colors and surface brightness), the remaining factor of
the overrepresentation of TDEs in those rare subclass of galaxies is
${\sim}$25--48 \citep{LawSmith17}. The underlying mechanism(s) responsible for
this preference is still unclear. Some proposed explanations include binary MBH
formation due to a recent galaxy merger, stellar orbital perturbations induced
by what has also caused the starburst in the past, a unique population of stars
such as an evolved population of A giants which are susceptible for disruptions,
and high central stellar density distributions or concentrations (e.g., see
\citealt{Alexander17, LawSmith17, French20, Gezari21}).  To properly interpret
these observations and thereby distinguish different dynamical mechanisms
occurring in the galactic nuclei to produce TDEs, it is necessary to have a
thorough understanding of these mechanisms, especially on the dependence of
those theoretically predicted rates on the properties of the central MBHs and
their host galaxies.

In a spherical stellar system composed of a central MBH and surrounding stars,
the tidal radius for a given type of stars and the MBH event horizon determine
the size of the loss cone in the phase space of the stellar energy and angular
momentum, in the sense that a star at a given energy can be consumed (either
tidally disrupted or swallowed as a whole) by the MBH when its angular momentum
falls below a critical value. The rate of stellar consumptions is determined by
the refilling rate of low-angular-momentum stars into the loss cone, since stars
initially in the loss cone will be quickly consumed (e.g., within an orbital
period). The two-body relaxation process of stars can set a lower limit for the
refilling rate of low-angular-momentum stars into the loss cone.  In realistic
stellar systems, some other mechanisms may also work or even play a dominant
role in refilling the loss cone, and therefore enhance the stellar consumption
rates. Possible mechanisms include the resonant relaxation \citep{RT96,
Hopman06RR}, massive perturbers which may accelerate the two-body relaxation
process \citep{Perets07}, binary MBHs or recoiled MBHs \citep{Ivanov05,
CSMetal11, Stone11}, as well as the stellar orbital precession within
nonspherical galaxy gravitational potentials (MT99; \citealt{Yu02, Vasiliev14}).

In an earlier work (\citealt{CYL20tde}, hereafter CYL20), a statistical study is
conducted on the cosmic distributions of stellar tidal disruptions by MBHs at
galactic centers, due to the combined effects of two-body relaxation and orbital
precession of stars in triaxial galaxy potentials; and the statistical results
reveal the correlations of the stellar consumption rate with both the MBH mass
$M\bh$ and the inner slope of the galaxy surface brightness profile $\gamma$. A
negative correlation between the stellar consumption rate (per galaxy) with the
MBH mass (e.g., \citealt{Wang04,Stone16}) is found to exist only for
$M\bh\lesssim 10^7\msun$, and the correlation becomes positive for $M\bh\gtrsim
10^7\msun$. At a given MBH mass $M\bh$, the stellar consumption rate is higher
in galaxies with larger $\gamma$ (steeper inner surface brightness profile) if
$M\bh\lesssim 10^7\msun$, but insensitive to $\gamma$ for MBHs with larger
masses. In triaxial galaxy potentials, the phase space of the loss cone
described for spherical galaxies can be replaced by the phase space of a general
loss region to incorporate the stars that can precess onto low angular momentum
orbits. 
The dichotomic trends of the stellar consumption rate at different
MBH mass ranges can be explained by that the dominant fluxes of stellar
consumption have different origins of the stellar low-angular-momentum orbits.

A further quantitative understanding of the origin of these correlations is of
great importance as it provides the key to distinguish the dominant mechanism(s)
of TDEs under different circumstances. Apart from that, the current and future
TDE observations shed new light on the demographic studies of the MBH
population, especially the mass function and the occupation fraction of MBHs at
the low-mass end (e.g., \citealt{Stone16, Fialkov17}); the correlations of the
stellar consumption with the MBH/galaxy properties cannot be ignored in a proper
interpretation of the observational results, and should be accompanied with a
quantitative understanding of the origin. To understand the origin of these
correlations, we employ an analytical model considering power-law stellar
distribution under the Keplerian potential of the central MBH. We first verify
that this simplified model provides a fairly good approximation when evaluating
the event rate of stellar consumption due to the two mechanisms, i.e., two-body
relaxation and stellar orbital precession in nonspherical potentials. Then we
identify the dominant factor(s) responsible for the slope and the scatter of
each correlation. 

The paper is organized as follows. In Section~\ref{sec:misc}, we first briefly
describe the galaxy sample used in this study. Then with the sample galaxies, we
show the correlations of the stellar consumption rates with the MBH mass and the
galaxy inner stellar distribution, as obtained in CYL20. In
Section~\ref{sec:model}, we construct the analytical model and obtain the
approximated expression for the stellar consumption rate due to either mechanism
and inspect the contribution from different terms in the approximated
expressions to these correlations. 
In Section~\ref{sec:discuss}, we
explore the possibility of using such correlations to explain the observed
overrepresentation of TDEs in those rare E+A/poststarburst galaxies. 
Based on the quantitative correlations constructed from the analytical model, we also generalize the discussion from the consumption rates of solar-type stars to those of other different types of stars.

\section{Galaxy sample and rate correlations}
\label{sec:misc}

\subsection{Galaxy sample}
\label{sec:misc:data}

In this study, we adopt the two observational samples of early-type galaxies
given by \citet{Lauer07sb} and \citet{Krajnovic13} to investigate the
correlation between the stellar consumption rate by the central MBH and either
the MBH mass or the host galaxy inner stellar distribution. For both samples,
high-spatial-resolution observations of the galaxy surface brightness by $HST$
are available, which are crucial for TDE studies since stars from the inner
region of the host galaxy (e.g., inside the influential radius of the central
MBH) contribute significantly to the stellar consumption (MT99). Note that these
two galaxy samples have also been adopted by CYL20 to study the cosmic
distributions of the stellar tidal disruptions and by \citet{CYL20bbh} to study
the properties of the cosmic population of binary MBHs as well as their
gravitational wave emissions.

For those early-type galaxies, their surface brightness profiles $I(R)$ can be
well described by the Nuker law \citep{Lauer95}. The best-fitting Nuker-law
parameters for the sample galaxies can be found in each source paper. When
calculating the stellar disruption/consumption rate, the mass-to-light ratio
$M/L$ of the galaxy is also required, since $M/L$ and the surface brightness
profile together determine the mass density distribution of the galaxy if it is
spherically distributed. For galaxies in \citet{Lauer07sb}, we obtain their
$M/L$  following Section~2.1 of \citet{Stone16}, i.e., their Equations~(4)-(5).
For galaxies in \citet{Krajnovic13}, we note that their $r$-band $M/L$ can be
found in \citet{Cappellari13}, but their Nuker-law parameters were fitted using
the $V$-band observations. In this study, we assume that the $V$-band $M/L$ for
these galaxies are the same as their $r$-band values. We estimate the MBH masses
$M\bh$ for the sample galaxies based on empirical scaling relations. For
galaxies in \citet{Lauer07sb}, we follow Section~2.1 of \citet{Stone16}, i.e.,
adopting the $M\bh$--$\sigma\rme$ relation from \citet{MM13} when $\sigma\rme$
is available in \citet{Lauer07bh}, while adopting the $M\bh$--$L_V$ relation
from \citet{MM13} otherwise. For galaxies in \citet{Krajnovic13}, we adopt the
$M\bh$--$\sigma\rme$ relation from \citet{MM13}, with $\sigma\rme$ taken from
\citet{Cappellari13}. When studying the stellar consumption due to loss-region
draining, we assume that each sample galaxy has a triaxial shape characterized
by $p_\rho=0.9$ and $q_\rho=0.8$, where $p_\rho$ and $q_\rho$ represent the
medium-to-major and minor-to-major axis ratios of the galaxy mass density
distribution.\footnote{We adopt this shape ($p_\rho=0.9, q_\rho=0.8$) as a
representative of generic triaxial stellar systems. The corresponding
triaxiality parameter is $T_\rho=(1-p_\rho^2)/(1-q_\rho^2)\simeq 0.53$. As seen
from Figure~7 of \citet{CYL20bbh}, the loss region (i.e., the stellar reservoir
for the refilling of the loss cone in triaxial systems) changes abruptly when
$T_\rho$ is very close to $0/1$ or $q_\rho$ is very close to $1$, while it
changes little with other $T_\rho$ and $q_\rho$ values. Therefore, the results
obtained in this study will hold even when the host galaxy has a different
shape, as long as the shape is not very close to perfectly spherical or
axisymmetric ones.}

\subsection{Stellar consumption rate correlations}
\label{sec:misc:corr}

In a spherical stellar system, the rate of stellar consumption by the central
MBH due to two-body relaxation can be evaluated by solving the Fokker-Planck
equation and the solution can be found in MT99 (see also \citealt{LS77, CK78}),
i.e.,
\begin{equation}
\Flc(\calE)d\calE = 
\frac{4\pi^2 \bar{f}(\calE)P(\calE)\bar{\mu}(\calE)J\rmc^2(\calE)d\calE}
{\ln R_0^{-1}(\calE)}.
\label{eq:Flc}
\end{equation}
Here $\bar{f}(\calE)$ is the ``isotropized'' distribution function (see Eq.~21
of MT99), $P(\calE)\equiv P(\calE,J^2=0)$ is the orbital period of a test
particle with ``specific binding energy'' $\calE$ (hereafter abbreviated as
energy) and zero ``specific angular momentum'' $J$ (hereafter abbreviated as
angular momentum), $\bar{\mu}(\calE)$ is the orbital-averaged diffusion
coefficient of the dimensionless angular momentum $R(\calE)\equiv
J^2(\calE)/J\rmc^2(\calE)$ in the limit $R\rightarrow 0$, with $J\rmc(\calE)$
denoting the angular momentum of a star on a circular orbit with energy $\calE$,
$R_0(\calE)$ marks the value of $R(\calE)$ at which the distribution function
falls to zero. $R_0(\calE)$ is given by
\begin{equation}
R_0(\calE) = R\lc(\calE)\times 
\left\lbrace \begin{split}
&\exp(-q) & q \ge 1 \\
& \exp(-0.186q-0.824\sqrt{q}) & q<1
\end{split}, \right.
\label{eq:R0}
\end{equation}
where $R\lc(\calE)\equiv J^2\lc(\calE)/J^2\rmc(\calE)$ represents the relative
size of the loss cone at energy $\calE$ in the phase space, with $J\lc$ denoting
the loss-cone angular momentum and $q(\calE)\equiv P(\calE)\bar{\mu}(\calE)/
R\lc(\calE)$ representing the ratio of the change of $R(\calE)$ for radial
orbits during an orbital period to $R\lc(\calE)$ (MT99).

\appendblock{
The stellar consumption rate due to loss-region draining in a nonspherical
potential at a sufficiently long consumption time $T$ can be approximated by
\begin{equation}
\Fdrain(\calE)d\calE = 4\pi^2 f(\calE) J\lc^2(\calE)
\exp\left[-\frac{T}{P(\calE)}\frac{J\lc^2(\calE)}{J\lr^2(\calE)}\right]d\calE,
\label{eq:Fdrain}
\end{equation}
where in triaxial galaxies $J\lr(\calE)$ corresponds to the characteristic angular momentum at energy
$\calE$ below which stars can precess into the loss cone, and it is determined
by the nonspherical shape of the host galaxy (see Eq.~9 in CYL20 and Eq.~52
in MT99). 
Due to the draining, stars initially inside the loss region are gradually
depleted, and stars initially outside the loss region can be diffused into it
due to two-body relaxation. As shown in CYL20, in triaxial galaxies, the
loss-region refilling rate can be obtained by generalizing the analysis done for
spherical systems, i.e., by replacing $J\lc$ in Equations
\eqref{eq:Flc}--\eqref{eq:R0} with the angular momentum $J\lr$ characterizing
the size of the loss region. 
When the stars initially in the loss region are depleted significantly at
$T\gg P(\calE){J\lr^2(\calE)}/{J\lc^2(\calE)}$,  the loss-region refilling
rate due to two-body relaxation can be larger than the loss-region draining
rate given by Equation (\ref{eq:Fdrain}) and thus dominate the stellar consumption rate.
}

When the loss-region refilling rate due to two-body relaxation dominates
the stellar consumption rate, compared with the rate obtained through two-body
relaxation in spherical systems, the correction of the stellar consumption rates
for triaxial galaxies can be mainly due to the change of the logarithm term of
$\ln R_0^{-1}(\calE)$ in Equations \eqref{eq:Flc}--\eqref{eq:R0}; and compared
to the estimates for spherical systems, the average stellar consumption rates in
triaxial galaxies are increased roughly by a similar factor of $f^{\rm tri}\sim$3--5 for
low-mass MBHs (with masses ranging from $M\bh\sim 10^5$--$10^7M_\odot$;
see Figure 3 or 4 in CYL20). 
This rate increasing factor is almost independent of the MBH mass for low-mass
MBHs (when $T\gtrsim 1$\gyr; see Fig.~5 of CYL20).  That is, the
consumption rate--MBH mass correlation obtained by only considering two-body
relaxation in spherical systems is similar to that by considering
loss-region draining and refilling, except for a normalization difference by a
factor of $f^{\rm tri}\sim 3-5$  (see Fig.~3 of CYL20). 
Thus, below for the purpose of this work on discussing the changing tendencies
of the stellar consumption rates with properties of MBHs and host galaxies and
for simplicity, we use Equations~\eqref{eq:Flc}--\eqref{eq:R0} obtained due to
the two-body relaxation for spherical systems to analyze parts of the origin of
the correlations.

\begin{figure*}[!htb]
\centering
\includegraphics[width=0.95\textwidth]{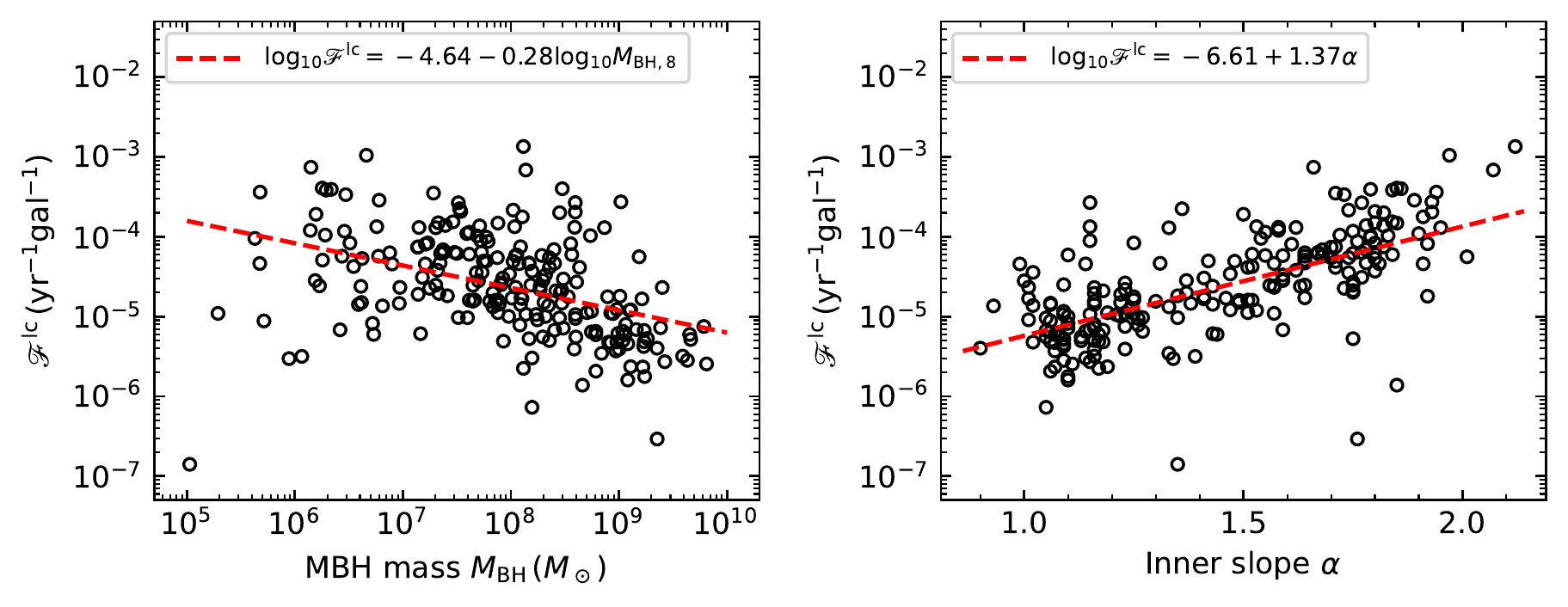}
\caption{Dependence of the stellar consumption rate by the central MBH due to
two-body relaxation $\scrFlc$ on the mass of the central MBH $M\bh$ (left panel)
and on the inner slope of the galaxy stellar number/mass density distribution
$\alpha$ (right panel). Each open circle represents a galaxy in the sample. The
red dashed lines in the left and the right panels represent the best fits to the
correlation between $\scrFlc$ and $M\bh$ and the correlation between $\scrFlc$
and $\alpha$, respectively. The best fit form is obtained by using the linear
least square regression algorithm and is also labeled in each panel. This figure
shows the slopes and scatters of both the correlation between $\scrFlc$ and
$M\bh$ and that between $\scrFlc$ and $\alpha$.}
\label{fig:f1}
\end{figure*}

\begin{figure*}[!htb]
\centering
\includegraphics[width=0.95\textwidth]{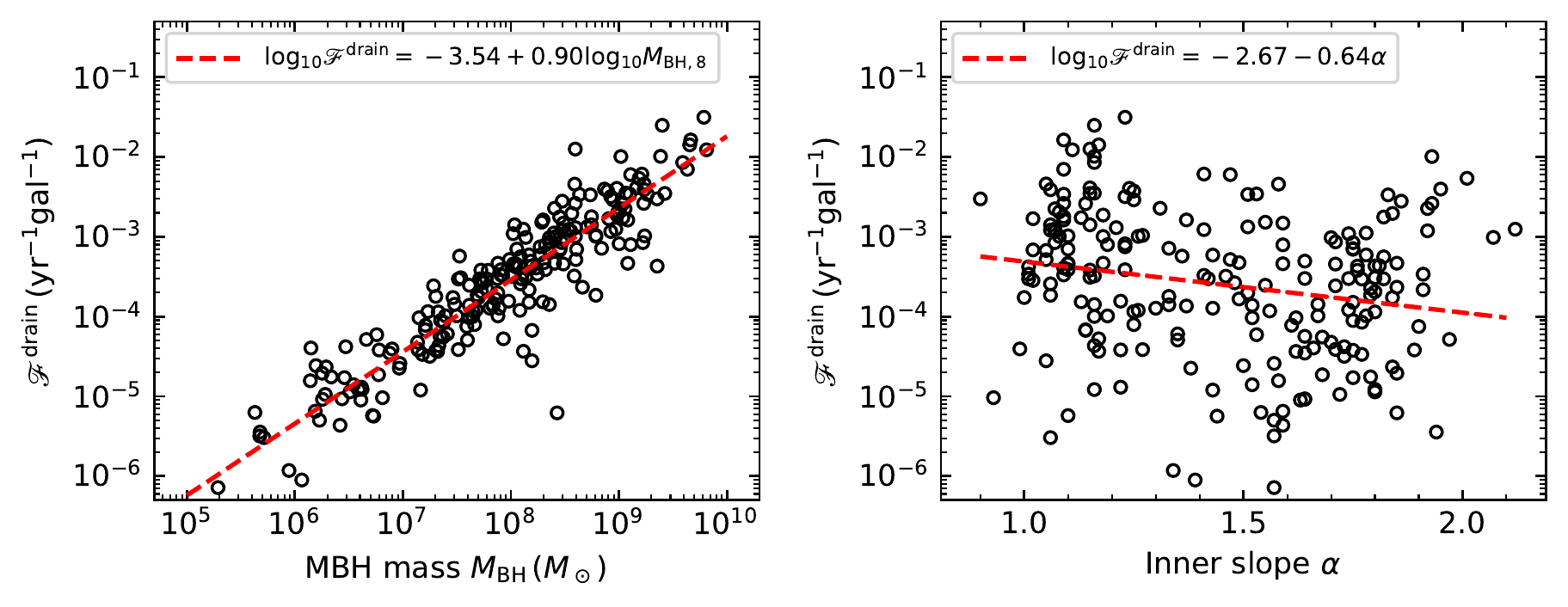}
\caption{Legends are the same as those for Fig.~\ref{fig:f1}, except that the
stellar consumption is due to the draining of the loss region in nonspherical
potentials. This figure shows the positive correlation between $\scrFdrain$ and
$M\bh$, and the mildly negative correlation between $\scrFdrain$ and $\alpha$.}
\label{fig:f2}
\end{figure*}

The origin of the correlations of the stellar consumption rates with properties
of MBHs and host galaxies can be analyzed through the functions of $\scrFlc =
\int\Flc(\calE)d\calE$ or $\scrFdrain = \int\Fdrain(\calE)d\calE$, where
$\scrFlc$ is the total rate of stellar consumption due to two-body relaxation
obtained by assuming that galaxies are spherical, and $\scrFdrain$ is the total
rate of stellar consumption due to the draining of the loss region obtained by
assuming that galaxies are non-spherical. 
By taking into account that at some $\calE$, $\Fdrain(\calE)$ can become subdominant compared to the
corresponding loss region refilling rate due to two-body relaxation, the stellar consumption rate (obtained by the integration over $\calE$) in a triaxial galaxy, $\sim\int\max[f^{\rm tri}\Flc(\calE),\Fdrain(\calE)]d\calE$, is expected to be larger than 
$\max(f^{\rm tri}\scrFlc,\scrFdrain)$, as well as being lower than
the sum of $f^{\rm tri}\scrFlc+\scrFdrain$.
Thus the stellar consumption rate is expected to be $\sim\scrFdrain$
if $\scrFdrain\gg f^{\rm tri}\scrFlc$,  and $\sim f^{\rm tri}\scrFlc$
if $\scrFdrain\ll f^{\rm tri}\scrFlc$;
 As to be seen below (e.g., Figs.~\ref{fig:f1}-\ref{fig:f2}), 
 both of those two cases cover a significantly large parameter space of host galaxy properties,
so that the functions of $\scrFlc$ and $\scrFdrain$ can be used to analyze the origin of the correlations of the stellar consumption rates. 
We estimate those stellar consumption rates of $\scrFlc$ and $\scrFdrain$ for each sample galaxy according to
Equations~\eqref{eq:Flc}--\eqref{eq:Fdrain}. We refer to CYL20 for some detailed
procedures to calculate the relevant quantities in these equations. 

Figure~\ref{fig:f1} shows the dependence of the stellar consumption rate due to
two-body relaxation $\scrFlc$ on the MBH mass $M\bh$ (left panel) and on the
inner slope of the galaxy stellar number/mass density distribution $\alpha$
(right panel), respectively. The stellar consumption rate of each galaxy is
evaluated based on Equation~\eqref{eq:Flc}. As seen from this figure, $\scrFlc$
has a negative correlation with $M\bh$, and a positive one with $\alpha$. We
conduct linear fittings to the $\log\scrFlc-\log M\bh$ relation (left panel) and
the $\log\scrFlc-\alpha$ relation (right panel), respectively. The best-fitting
relations are shown by the red dashed line in each panel, with the corresponding
slope and intercept labeled in the panel. Similarly, in Figure~\ref{fig:f2}, we
show the dependence of the stellar consumption rate due to loss-region draining
in nonspherical potentials $\scrFdrain$ on the mass of the central MBH $M\bh$
and on the inner slope of the galaxy stellar number/mass density distribution
$\alpha$, where we set $T=10\gyr$. In contrast to the case of
two-body relaxation, $\scrFdrain$ has a positive correlation with $M\bh$ and a
mildly negative correlation with $\alpha$. The best-fitting relations are also
shown by the red dashed lines in both panels.

As seen from Figures~\ref{fig:f1} and \ref{fig:f2}, we have $\scrFlc\ga\scrFdrain$ at
$M\bh\lesssim 10^7\msun$ and $\scrFlc\la\scrFdrain$ at $M\bh\gtrsim 10^7\msun$.
Correspondingly, we apply the correlations of $\scrFlc$ and $\scrFdrain$ to
interpret the correlation tendencies of the stellar consumption rates in those
low-mass and high-mass ranges of MBHs, respectively.

Note that the fits shown in Figures~\ref{fig:f1} and \ref{fig:f2} cover the
whole MBH mass range, as it is a clear way to show the effects of the same
mechanism through a large mass range. If the fit is limited to only a small mass
range of $M\bh\lesssim 10^7\msun$, specifically the correlation between
$\scrFlc$ and $M\bh$ could appear quite mild or may not be negative due to the
small mass range and large rate scatters. In general, the other correlation
tendencies are found not to be affected significantly even if limiting the fit
mass range to either $M\bh\lesssim 10^7\msun$ or $M\bh\gtrsim 10^7\msun$.  In
this work, for connecting the stellar consumption rates with the underlying
mechanisms and for simplicity, we use the fit results of the whole mass range
below.

\section{The power-law stellar distribution model}
\label{sec:model}

\appendblock{
\begin{deluxetable*}{LRRcRR}
\tablecaption{Best-fit parameters and their $1\sigma$ uncertainties for the
different correlations shown in Figures~\ref{fig:f1}--\ref{fig:f2}, \ref{fig:f4}, and \ref{fig:f6}. \label{tab:t1}}
\tablehead{ & \multicolumn{2}{c}{$\log_{10}Y=b_M+k_M\log_{10}M_{\rm BH,8}$} &&
\multicolumn{2}{c}{$\log_{10}Y=b_\alpha+k_\alpha\alpha$} \\
\cline{2-3} \cline{5-6}
\dcolhead{Y} & \dcolhead{b_M} & \dcolhead{k_M} && \dcolhead{b_\alpha} &
\dcolhead{k_\alpha}}
\startdata
\scrFlc              & -4.64(0.04) & -0.28(0.04) && -6.61(0.18) &  1.37(0.12) \\
\mbhsigmarelaxunit   & -0.40(0.04) & -0.47(0.04) && -2.35(0.20) &  1.37(0.13) \\
\zeta(\alpha)        & -0.11(0.01) & -0.06(0.01) && -1.02(0.01) &  0.63(0.00) \\
\phi(\omega)         & -0.72(0.02) &  0.26(0.02) &&  0.57(0.11) & -0.90(0.07) \\
\hline
\scrFdrain           & -3.54(0.03) &  0.90(0.03) && -2.67(0.31) & -0.64(0.21) \\
\chi(\alpha)         &  0.26(0.01) &  0.16(0.01) &&  1.18(0.04) & -0.64(0.03) \\
\mbhsigmadrain       & -0.37(0.04) &  0.69(0.04) && -0.71(0.29) &  0.20(0.19) \\
\Rlrterm             & -0.18(0.01) &  0.01(0.01) &&  0.12(0.02) & -0.21(0.02)
\enddata
\end{deluxetable*}
}

\begin{figure}[!htb]
\centering
\includegraphics[scale=0.8]{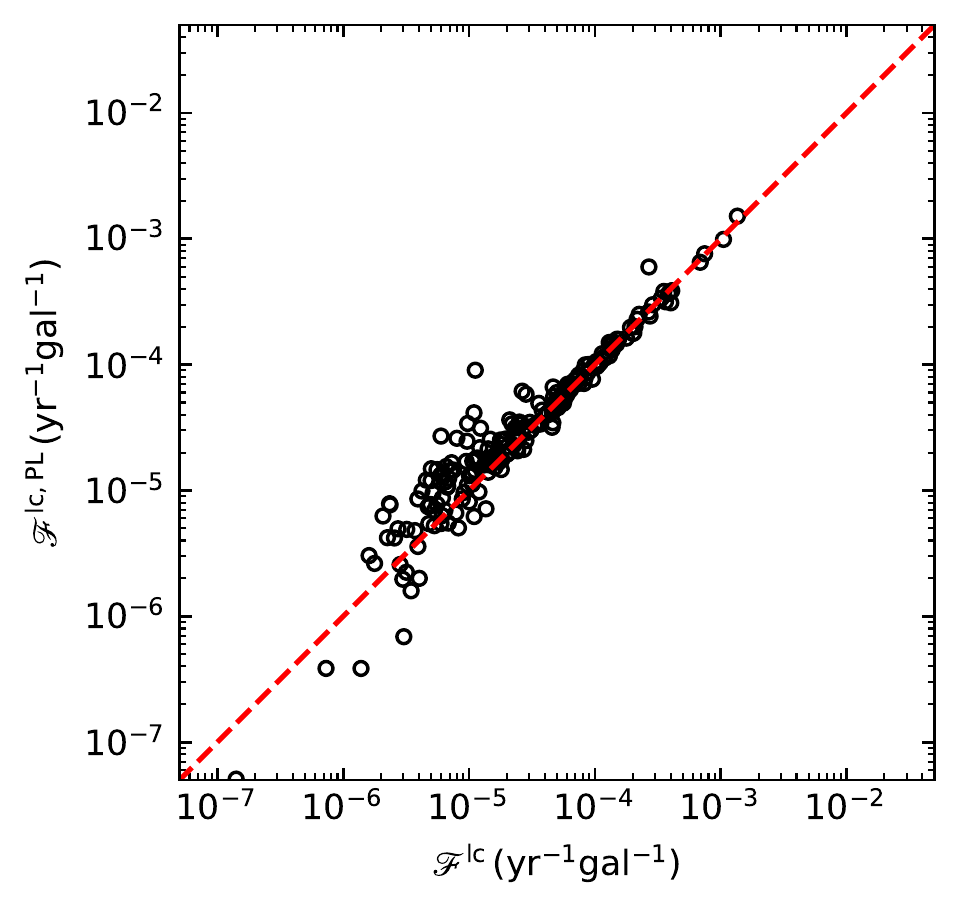}
\caption{Comparisons of the calculated stellar consumption rates due to the
two-body relaxation and those obtained approximately by adopting the simple
power-law model described in Section~\ref{sec:model}, where the galaxies are
assumed to be spherical. This figure verifies that the simple power-law model is
effective in estimating the stellar consumption rates by the central MBHs due to
the two-body relaxation, and therefore can be used to inspect the correlations
of the stellar consumption rate with both the MBH mass and the inner stellar
distribution of the galaxy.}
\label{fig:f3}
\end{figure}

The explanation to the correlations shown in Figures~\ref{fig:f1} and
\ref{fig:f2} obtained by using Equations~\eqref{eq:Flc}-\eqref{eq:Fdrain} is not
explicit, since many of the terms in these equations are calculated numerically
for individual galaxies. In the following subsections, we employ a power-law
model to obtain approximate expressions for the stellar consumption rates due to
both the mechanisms. By saying ``power-law model'' we mean that the stellar
number/mass density distribution of the galaxy can be described by a single
power law and only the Keplerian potential of the central MBH is considered. We
investigate the relative contributions of different terms in this simplified
model to the stellar consumption rate and thus figure out the dominant factors
that lead to these two correlations. As a counterpart, we define the ``full
model'' as the one evaluating the stellar consumption rates through
Equations~\eqref{eq:Flc}--\eqref{eq:Fdrain}. Below, we describe the details of
the power-law model.

In the simplified model, we assume that the distribution of stars around the
central MBH follows a single power law, i.e., $n_\ast(r)\propto r^{-\alpha}$
where $n_\ast(r)$ is the spatial number density of stars at radius $r$ and
$\alpha$ is the power-law index. We define the influential radius of the MBH
$r\rmh$ to be the radius within which the enclosed stellar mass equals the MBH
mass, i.e., $M\rmh\equiv M_\ast(r\leq r\rmh)= M\bh$. We make the power-law
assumption based on the finding that stellar diffusion in the energy space
gradually drives the distribution of stars in a system containing a central MBH
towards a power-law cusp \citep{BW76}. In practice, we adopt the Nuker law
\citep{Lauer95} as a generic description of the surface brightness profiles of
the galaxies. The break radii of realistic galaxies $R\rmb$ are typically much
larger than the influential radii of the MBHs $r\rmh$, ensuring that the
power-law density distribution a fairly good approximation for stars inside
$r\rmh$. Therefore, the spatial number density of stars surrounding the MBH
could be formulated as
\begin{equation}
n_\ast(r) = \frac{(3-\alpha)N\rmh}{4\pi r\rmh^3}
\left({\frac{r}{r\rmh}}\right)^{-\alpha},
\label{eq:nstar}
\end{equation}
where $N\rmh$ is the total number of stars enclosed in $r\rmh$. The total number
of stars enclosed in any given radius $r$ is $N_\ast(\leq r)=
N\rmh(r/r\rmh)^{3-\alpha}$. If we assume identical mass of stars, we also have
$M_\ast(\leq r)= M\rmh(r/r\rmh)^{3-\alpha}$, with $M_\ast(\leq r)$ being the
total stellar mass enclosed in radius $r$. Throughout this paper, we make the
approximation that $\alpha=\gamma+1$, where $\gamma$ is the inner slope of the
Nuker-law surface brightness profile. It turns out to be a good approximation
for sample galaxies with $\gamma\ga 0.2$. For galaxies with $\gamma \la 0.2$,
the approximation tends to overestimate the inner slope of the host galaxy
stellar number/mass density profile. However, the main conclusions obtained in
this paper do not change by a small number of such galaxies in the sample.

In the power-law model, we ignore the self-gravity of the stellar system and
assume that the potential is mainly contributed by the central MBH. We make this
assumption based on the fact that the loss-cone consumption rate due to either
mechanism peaks around the influential radius of the central MBH if we convert
energy to radius based on the radial energy profile for stars in circular
orbits, i.e., $\calE(r)$. For example, for most of the sample galaxies,
$\calE\Flc(\calE)$ peaks around $r\rmh$ within a factor of $2$, while
$\calE\Fdrain(\calE)$ peaks around $r\rmh$ within a factor of $3$ when the
consumption time is fixed to $T=10\gyr$. Adopting a different $T$ does not
significantly affect our results. Under the Keplerian potential of the central
MBH, we evaluate the ergodic distribution function based on the Eddington's
formula. The resulting differential energy distribution of stars
$N(\calE)d\calE$ (i.e., the number of stars in the stellar system with energy in
the range of $\calE\rightarrow \calE+d\calE$) is expressed by
\begin{equation}
N(\calE)d\calE = 
\xi(\alpha)N\rmh\left(\frac{\calE}{\calE\rmh}\right)^{\alpha-3} d\ln\calE,
\label{eq:NcalE}
\end{equation}
where
\begin{equation}
\xi(\alpha) = \frac{\sqrt{\pi}}{8}(3-\alpha)2^{3-\alpha}
\frac{\Gamma(\alpha+1)}{\Gamma(\alpha-\frac{1}{2})},
\label{eq:xialpha}
\end{equation}
and $\calE\rmh\equiv \sigma\rmh^2/2$ where $\sigma\rmh^2\equiv GM\bh/r\rmh$.

\subsection{Two-body relaxation mechanism} 
\label{sec:model:lc}

\begin{figure*}[!htb]
\centering
\includegraphics[width=0.8\textwidth]{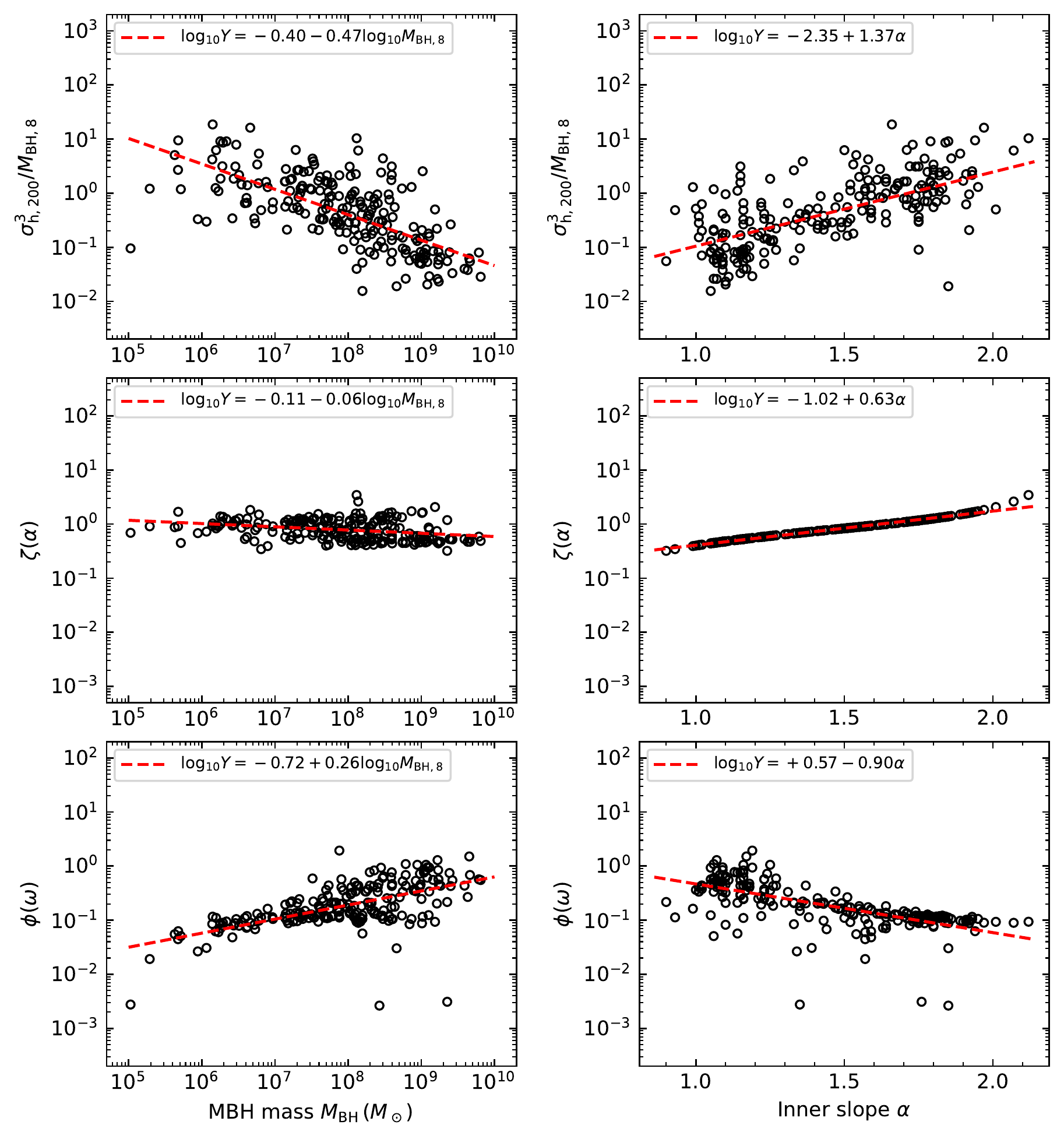}
\caption{Dependence of the terms $\sigma^3 M\bh^{-1}$, $\zeta(\alpha)$, and
$\phi(\omega)$ in the approximated stellar consumption rate due to the two-body
relaxation in the power-law model $\scrFlcPL$ (Eq.~\ref{eq:scrFlcPL}) on the MBH
mass $M\bh$ (left panels) and the inner slope of the host galaxy mass density
distribution $\alpha$ (right panels). From top to bottom, the three inspected
terms are $\mbhsigmarelax$, $\zeta(\alpha)$ (Eq.~\ref{eq:zetaalpha}), and
$\phi(\omega)$ (Eq.~\ref{eq:phiomega}), respectively. In the top two panels,
$\sigma_{\rm h,200}$ and $M_{\rm BH,8}$ are $\sigma\rmh$ and $M\bh$ in unit of
$200\kms$ and $10^8\msun$, respectively. The best-fit is shown by the red dashed
line in each panel, and the best fit form is also marked there with $Y$
representing the quantity of $\sigma_{\rm h,200}^3 M\bh^{-1}$, $\zeta(\alpha)$,
and $\phi(\omega)$ in the top, middle, and bottom panels, respectively. Note
that for the convenience of direct comparisons of the scatters with
Fig.~\ref{fig:f1}, all the panels here share the same decades along the $y$-axis
with the panels in Fig.~\ref{fig:f1}. This figure aims to reveal the dominant
contributors to the slopes and scatters of both the correlation between
$\scrFlc$ and $M\bh$ and that between $\scrFlc$ and $\alpha$. See
Section~\ref{sec:model:lc} for more details.}
\label{fig:f4}
\end{figure*}

We first consider the stellar consumption due to the two-body relaxation by
assuming the galaxy gravitational potential are spherical. When the Keplerian
potential of the central MBH dominates, we have $P(\calE,J^2) \simeq P(\calE)
\simeq 2\pi GM\bh/(2\calE)^{3/2}$; and the differential energy distribution can
be expressed as $N(\calE)\simeq 4\pi^2 f(\calE)P(\calE)J\rmc^2(\calE)$ (i.e.,
see Eq.~5 of MT99). In this case, Equation~\eqref{eq:Flc} can be approximated as
\begin{equation}
\Flc(\calE)d\calE 
\simeq \frac{N(\calE)\bar{\mu}(\calE) d\calE}{\ln R_0^{-1}(\calE)}
\simeq \frac{N(\calE)}{t\rmr(\calE)}\frac{d\calE}{\ln R_0^{-1}(\calE)},
\label{eq:FlcPL}
\end{equation}
where in the latter expression we have used the inverse of the relaxation
timescale $1/t\rmr(\calE)$ to approximate the orbital-averaged diffusion
coefficient of the angular momentum, $\bar{\mu}(\calE)$. The relaxation
timescale $t\rmr(\calE)$ can be expressed as (see \citealt{BarOr13} and Eq.~1 of
\citealt{Alexander17})
\begin{equation}
t\rmr(\calE) \simeq \eta\frac{Q^2P\rmc(\calE)}{N_\ast(\leq r)\ln Q}
\simeq \frac{2\pi \eta GM\bh Q^2}{N\rmh\sigma\rmh^3\ln Q}
\left(\frac{\calE}{\calE\rmh}\right)^{3/2-\alpha},
\label{eq:trelax}
\end{equation}
where $\eta\simeq 1/8$, $Q\equiv M\bh/\langle m_\star\rangle\simeq N\rmh$ with
$\langle m_\star\rangle$ representing the mean mass of stars, and $P\rmc(\calE)$
is the orbital period of a circular orbit at energy $\calE$. To get the
rightmost expression in Equation~\eqref{eq:trelax}, we assume the dominance of
the Keplerian potential of the central MBH and use $r/r\rmh\simeq
(\calE/\calE\rmh)^{-1}$. Substituting Equations~\eqref{eq:NcalE} and
\eqref{eq:trelax} into Equation~\eqref{eq:FlcPL}, we have
\begin{equation}
\calE \Flc(\calE)
\simeq \frac{\xi(\alpha)\sigma\rmh^3\ln Q}{2\pi\eta GM\bh\ln R_0^{-1}(\calE)}
\left(\frac{\calE}{\calE\rmh}\right)^{2\alpha-9/2}.
\label{eq:EFE}
\end{equation}
If we omit the variance of $\lnR(\calE)$, which is a mildly decreasing function
of $\calE$ in the range $\calE\ga\calE\rmh$, then $\calE\Flc(\calE)$ is a
decreasing function of $\calE$ as long as $\alpha<9/4$, as seen from
Equation~\eqref{eq:EFE}. If the inner stellar distribution is core-like, i.e.,
$\alpha$ is small, $\calE\Flc(\calE)$ decreases relatively sharply with
increasing $\calE$, whereas if the inner stellar distribution is cusp-like with
large $\alpha$, $\calE\Flc(\calE)$ decreases relatively mildly with increasing
$\calE$. 

In real galaxies, $\calE\Flc(\calE)$ peaks at $\calE\peak\equiv \omega\calE\rmh$
where $\omega\simeq 1$. For $\calE < \calE\peak$, the self-gravity from the
stellar system cannot be neglected and Equation~\eqref{eq:EFE} does not work any
more. As a matter of fact, $\calE\Flc(\calE)$ is an increasing function of
$\calE$ at $\calE\leq \calE\peak$, instead of a decreasing function as described
by Equation~\eqref{eq:EFE} for $\calE >\calE\peak$. Therefore, the stellar
consumption rate can be divided into two parts separated by $\calE\peak$. For
the inner part with $\calE\geq \calE\peak$, the rate can be approximated by
integrating $\Flc(\calE)$ over the range $\calE\geq \calE\peak$ using
Equation~\eqref{eq:EFE}. For those real galaxies in our sample, we find that the
fraction of the stellar consumption rate contributed by the inner part (i.e.,
with $\calE\geq \calE\peak$) spans $0.1\sim 0.8$, and this fraction tends to be
larger for a galaxy with a larger $\alpha$. Therefore, to compensate for the
fraction of the rate contributed by those stars with $\calE\leq \calE\peak$
which are omitted in the integration, we introduce a fudge factor $\fudge$ by
integrating $\Flc(\calE)$ over the range $\calE\geq \fudge\calE\peak=
\fudge\omega\calE\rmh$ using Equation~\eqref{eq:EFE}. The resulting stellar
consumption rate in the power-law model is given by 
\begin{eqnarray}
\scrFlcPL & \simeq &
\int_{\fudge\omega\calE\rmh}^{\infty} \calE\Flc(\calE) d\ln\calE 
\simeq  \frac{\zeta(\alpha)\sigma\rmh^3(\fudge\omega)^{2\alpha-\frac{9}{2}}\ln Q}
{2\pi\eta GM\bh\ln R_0^{-1}(\omega\calE\rmh)} \nonumber \\
& = & \zeta(\alpha) \phi(\omega) \frac{\sigma\rmh^3}{M\bh} \frac{\ln Q}{2\pi G\eta},
\label{eq:scrFlcPL}
\end{eqnarray}
where $\zeta(\alpha)$ and $\phi(\omega)$ are defined as
\begin{equation}
\zeta(\alpha) = \frac{2}{9-4\alpha}\cdot\xi(\alpha),
\label{eq:zetaalpha}
\end{equation}
and
\begin{equation}
\phi(\omega) = 
\frac{(\fudge\omega)^{2\alpha-\frac{9}{2}}}{\lnR(\omega\calE\rmh)}.
\label{eq:phiomega}
\end{equation}
We find that $\fudge=1/2$ works well empirically. Therefore, we set $\fudge=1/2$
in the following analysis if not stated specifically.

We compare the stellar consumption rates calculated from Equations
\eqref{eq:Flc}--\eqref{eq:R0} and those obtained approximately from the
power-law model, as shown in Figure~\ref{fig:f3}. Apparently, the rates
estimated by using the simple power-law model are well consistent with those
calculated from the full model (see Eqs.~\ref{eq:Flc}--\ref{eq:Fdrain} and
\citealt{CYL20tde}), which verifies the effectiveness of the simple power-law
model in obtaining the correct rate and suggests the validity of using this
simple model to inspect the correlations between $\scrFlc$ with both $M\bh$ and
$\alpha$. According to Equation~\eqref{eq:scrFlcPL}, both the correlation
between $\scrFlc$ and $M\bh$ and that between $\scrFlc$ and $\alpha$ are
controlled by three terms, i.e., $\mbhsigmarelax$, $\zeta(\alpha)$, and
$\phi(\omega)$. Figure~\ref{fig:f4} shows the dependence of the above three
terms on the MBH mass $M\bh$ (left panels) and the inner slope of the host
galaxy mass density distribution $\alpha$ (right panels), separately. As seen
from this figure, $\sigma\rmh^3M\bh^{-1}$ negatively correlates with $M\bh$,
$\phi(\omega)$ positively correlates with $M\bh$, while the term $\zeta(\alpha)$
only weakly correlate with $M\bh$; both $\sigma\rmh^3 M\bh^{-1}$ and
$\zeta(\alpha)$ positively correlate with $\alpha$, while $\phi(\omega)$
negatively correlates with $\alpha$, To describe the relative contribution of
each term to the $\scrFlc-M\bh$ relation and the $\scrFlc-\alpha$ relation
quantitatively, we conduct a linear least square fitting to the data points
shown in each panel of Figure~\ref{fig:f4}. The best fit is shown by the red
dashed line in each panel and its form is also labeled there. The best-fit
parameters with their $1\sigma$ uncertainties are also listed in
Table~\ref{tab:t1}.

We identify the main factors that lead to the $\scrFlc-M\bh$ relation and the
$\scrFlc-\alpha$ relation quantitatively based on the fitting results (see Tab.~\ref{tab:t1}
and legends in Figs.~\ref{fig:f1} and \ref{fig:f4}).
\begin{itemize}
\item
For the $\scrFlc-M\bh$ relation, the best fit gives a slope of $-0.28$ (left
panel of Fig.~\ref{fig:f1}). The best fits shown in the left panels of
Figure~\ref{fig:f4} (from top to bottom) have slopes of $-0.47$, $-0.06$, and
$0.26$, respectively. The sum of these three slopes is $-0.27$, which can
approximately account for that found in the left panel of Figure~\ref{fig:f1}.
Among the three terms, $\mbhsigmarelax$ dominates the total slope. The
contribution from $\phi(\omega)$ can only cancel about half the amount of
$\mbhsigmarelax$, and $\zeta(\alpha)$ contributes the least to the total
slope. Apart from the slope, the data points in the top left panel of
Figure~\ref{fig:f4} have comparable scatters as those in the left panel of
Figure~\ref{fig:f1}, and are substantially larger than those in the middle left
or the bottom left panel of Figure~\ref{fig:f4}. Therefore, the intrinsic
scatter of the $\scrFlc-M\bh$ relation is also dominated by term
$\mbhsigmarelax$, i.e., by the intrinsic scatter of the correlation between
$\mbhsigmarelax$ and $M\bh$, which is in turn determined by the scatter of the
correlation between $\sigma\rmh$ and $M\bh$.
\item
For the $\scrFlc-\alpha$ relation, the best fit gives a slope of $1.37$ (right
panel of Fig.~\ref{fig:f1}). The best fits shown in the right panels of
Figure~\ref{fig:f4} (from top to bottom) give slopes of $1.37$, $0.63$, and
$-0.90$, respectively. The sum of these three slopes is $1.10$, which is close
to that found for the $\scrFlc-\alpha$ relation (right panel of
Fig.~\ref{fig:f1}), again suggests that the correlation between $\scrFlc$ and
$\alpha$ can be explained approximately by the combined effect of these three
terms. Among the three terms, $\mbhsigmarelax$ dominates the contribution to the
total slope. While $\zeta(\alpha)$ and $\phi(\omega)$ contribute to the total
slope in close amounts but different signs so as to be cancelled much. As seen also from the right panels
of Figure~\ref{fig:f4}, the scatters among the data points in the top panel
dominate over those in the middle or bottom panels. Therefore, we can conclude
that both the slope and the scatter of the correlation between $\scrFlc$ and
$\alpha$ are dominated by the correlation between $\mbhsigmarelax$ and $\alpha$.
\end{itemize}

\subsection{Loss-region draining in nonspherical potentials}
\label{sec:model:drain}

\begin{figure}[!htb]
\centering
\includegraphics[scale=0.8]{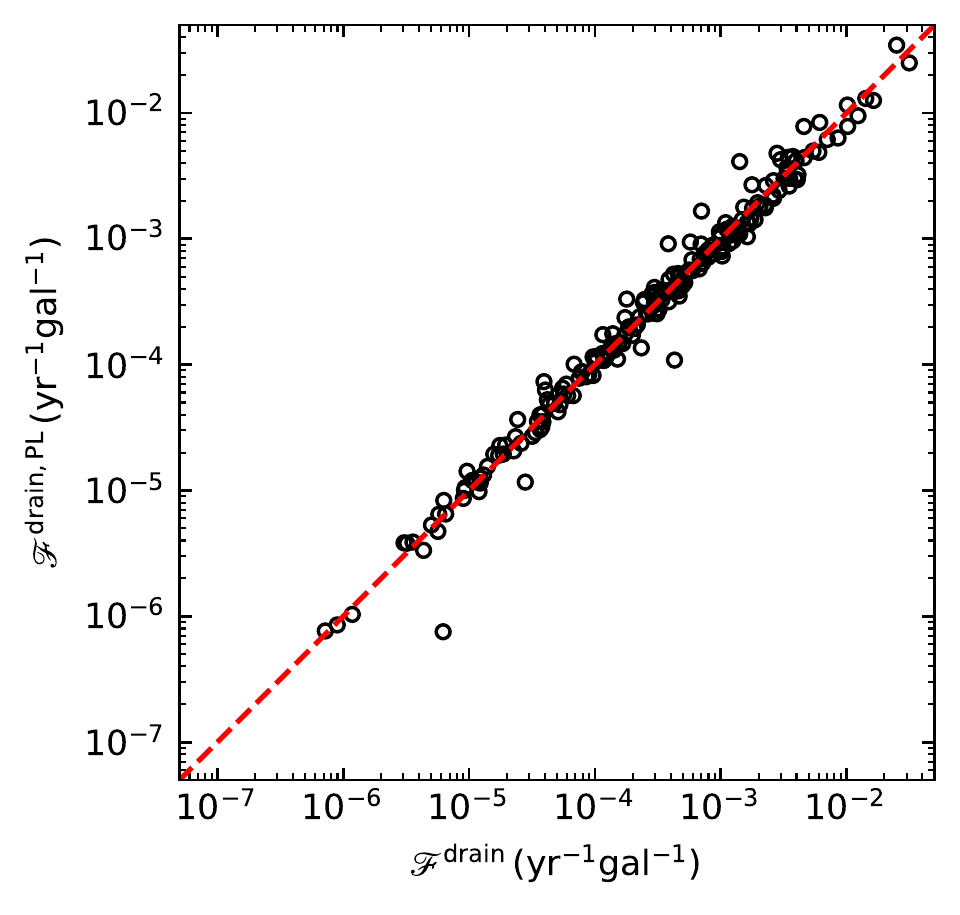}
\caption{Legends are similar to that for Fig.~\ref{fig:f3}, but the stellar
consumption rates are due to the draining of loss region in nonspherical
potentials.}
\label{fig:f5}
\end{figure}

\begin{figure*}[!htb]
\centering
\includegraphics[width=0.80\textwidth]{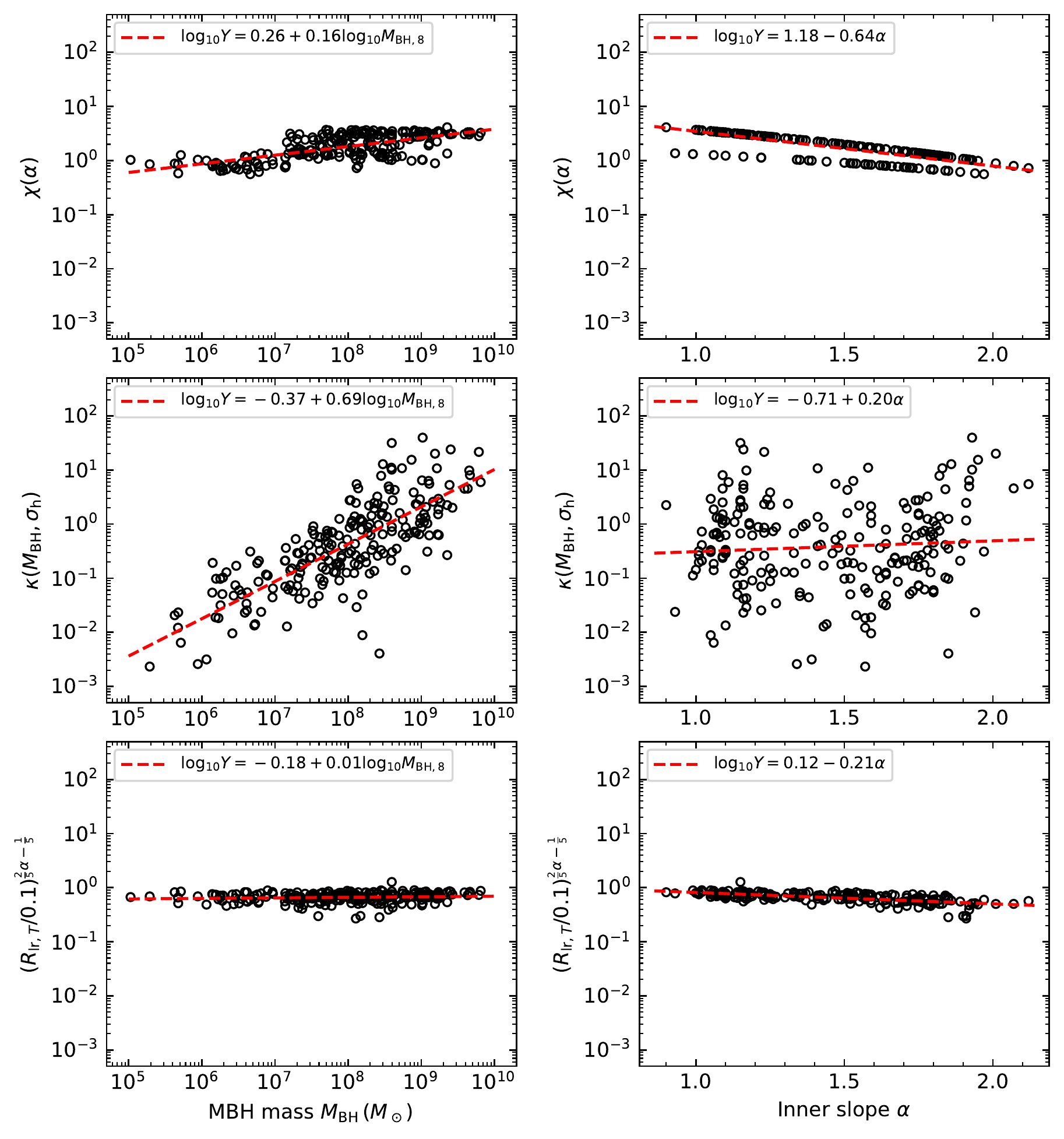}
\caption{Dependence of the terms in the approximated stellar consumption rate
due to draining of the loss region in nonspherical potentials in the power-law
model $\scrFdrainPL$ (Eqs.~\ref{eq:scrFdrainPL} and \ref{eq:scrFdrainPLunited})
on the MBH mass $M\bh$ (left panels) and the inner slope of the host galaxy mass
density distribution $\alpha$ (right panels). From top to bottom, the three
inspected terms are $\chi(\alpha)$ (Eq.~\ref{eq:chialpha}), $\mbhsigmadrain$
(Eq.~\ref{eq:mbhsigmadrain}), and $\Rlrterm$ (Eq.~\ref{eq:calET}), respectively.
The best fit to the data is shown by the red dashed line in each panel, and the
best fit form is also marked there with $Y$ representing $\chi(\alpha)$,
$\mbhsigmadrain$, and $\Rlrterm$, respectively. For the convenience of direct
comparisons of the scatters with Figure~\ref{fig:f2}, all the panels here share
the same decades along the $y$-axis with the panels in Figure~\ref{fig:f2}. This
figure aims to reveal the dominant contributors to the slopes and scatters of
the correlations either between $\scrFdrain$ and $M\bh$ or between $\scrFdrain$
and $\alpha$. See Section~\ref{sec:model:drain} for more details.}
\label{fig:f6}
\end{figure*}

We now consider the correlation between $\scrFdrain$ and $M\bh$ or $\alpha$ in
galaxies with nonspherical mass distributions. We define the loss-region
draining timescale $\tau(\calE)\equiv P(\calE) J\lr^2(\calE)/J\lc^2(\calE)$,
which characterizes how long the reservoir of loss-region stars at energy
$\calE$ can sustain the consumption by the central MBH. Similar as in the
loss-cone case, we define the dimensionless loss-region angular momentum
$R\lr(\calE)\equiv J\lr^2(\calE)/J\rmc^2(\calE)$, which characterizes the
relative size of the loss region at energy $\calE$ in the phase space. In
generic triaxial galaxies, $R\lr(\calE)$ can reach the order of $\sim$0.1 at
energies satisfying $\calE\la \calE\rmh$ and decreases with increasing $\calE$
for $\calE> \calE\rmh$. The draining timescale can be then expressed as
$\tau(\calE)= P(\calE) R\lr(\calE)/R\lc(\calE)$. Therefore, the flux of stars
being drained from the loss region into the loss cone to be consumed can be
expressed as
\begin{equation}
\Fdrain(\calE)d\calE \simeq
\frac{N(\calE)R\lr(\calE)d\calE}{\tau(\calE)}
\exp\left[-\frac{T}{\tau(\calE)}\right],
\label{eq:FdrainPL}
\end{equation}
where we have again made the approximation $P(\calE,J^2)\simeq P(\calE)\equiv
P(\calE,J^2=0)$. Assuming the dominance of the Keplerian potential of the
central MBH, we have $P(\calE)\simeq 2\pi GM\bh/(2\calE)^{3/2}$ and
$J\rmc(\calE)\simeq GM\bh/(2\calE)^{1/2}$. Define the consumption radius
$r\consp\equiv J^2\lc/2GM\bh$, then the loss-region draining timescale can be
approximated as
\begin{equation}
\tau(\calE)\simeq 
\frac{\pi (GM\bh)^2 R\lr(\calE)}{4\sqrt{2}\calE^{5/2}r\consp}.
\label{eq:taucalE}
\end{equation}
Since typically $\calE\Fdrain(\calE)$ is an increasing function of $\calE$, the
total stellar consumption rate at the draining time $T$ due to loss-region
draining is dominated by the flux of stars at $\calE\simeq \calE_T$, with
$\calE_T$ satisfies $\tau(\calE_T)=T$, which gives
\begin{equation}
\calE_T=\left[\frac{\pi(GM\bh)^2 R\lr{_{,T}}}{4\sqrt{2}T r\consp}\right]^{2/5},
\label{eq:calET}
\end{equation}
where $R\lrT = R\lr(\calE_T)$. At energy $\calE < \calE_T$ where the loss region
has not been exhausted yet, $\calE\Fdrain$ increases with $\calE$ in a manner
that $\calE\Fdrain \sim \calE^{\alpha-1/2}$ in the power-law model. Therefore,
the total stellar consumption rate due to the loss-region draining can be
evaluated as
\begin{eqnarray}
\scrFdrainPL & \simeq & 
\frac{2}{2\alpha-1}f_T\calE_T \Fdrain(\calE_T) \nonumber \\
& = & \frac{2\xi(\alpha)}{2\alpha-1}\frac{f_T N\rmh R\lrT}{T}
\left(\frac{\calE_T}{\calE\rmh}\right)^{\alpha-3},
\label{eq:scrFdrainPL}
\end{eqnarray}
where $\xi(\alpha)$ is given by Equation~\eqref{eq:xialpha}, $f_T$ is a fudge
factor introduced to make the rates estimated by
the power-law model consistent with those by the full model, and $f_T$ changes from $0.78$
at $T=0.1\gyr$ to $0.56$ at $T=10\gyr$. Since we analyze the rate correlations
at $T=10\gyr$, we adopt $f_T=0.56$ below.

The definition of the consumption radius $r\consp$ naturally leads to a
separatrix in the MBH mass $M\bh$, below which $r\consp = r\rmt= (f\rmt
M\bh/m_\star)^{1/3} R_\star$, with $m_\star$ and $R_\star$ being the mass and
radius of the star and $f\rmt$ being a dimensionless factor, while above which
$r\consp = 2r\swl= 8GM\bh/c^2$ ($r\swl\equiv 4GM\bh/c^2$; see Eq.~1 in CYL20).
For MBHs with masses being below the separatrix, setting $m_\star$ and $R_\star$
to be the solar mass and solar radius and $f\rmt=1$, we have
\begin{eqnarray}
& & \frac{\calE_T}{\calE\rmh} = \left[\frac{\pi (GM\bh)^2R\lrT} 
{T r\consp\sigma\rmh^5}\right]^{\frac{2}{5}} \nonumber \\
& & \simeq 1.24\left[\frac{M\bh}{10^8\msun}\right]^{\frac{2}{3}}
\left[\frac{\sigma\rmh}{200\kms}\right]^{-2}
\left[\frac{R\lrT}{0.1}\right]^{\frac{2}{5}}
\left[\frac{T}{10\gyr}\right]^{-\frac{2}{5}}. \nonumber \\
\label{eq:calET2calEhBelow}
\end{eqnarray}
Substituting Equation~\eqref{eq:calET2calEhBelow} into
Equation~\eqref{eq:scrFdrainPL}, the stellar consumption rate due to the
loss-region draining when the MBH mass is below the separatrix can be
approximated in the power-law model as
\begin{eqnarray}
\scrFdrainPL
\simeq 1.24^{\alpha-3}\frac{2\xi(\alpha)}{2\alpha-1} \times 10^{-3}f_T\pyr
\left[\frac{M\bh}{10^8\msun}\right]^{\frac{2\alpha-3}{3}}\nonumber \\
\left[\frac{\sigma\rmh}{200\kms}\right]^{6-2\alpha}
\left[\frac{R\lrT}{0.1}\right]^{\frac{2\alpha-1}{5}}
\left[\frac{10\gyr}{T}\right]^{\frac{2\alpha-1}{5}}.\nonumber \\
\label{eq:scrFdrainPLBelow}
\end{eqnarray}
On the other hand, when the MBHs have masses being above the separatrix, we have
\begin{eqnarray}
& & \frac{\calE_T}{\calE\rmh} 
= \left[\frac{\pi (GM\bh)^2R\lrT}{T r\consp\sigma\rmh^5}\right]^{\frac{2}{5}}
\simeq \nonumber \\ 
& &
0.735  \left[\frac{M\bh}{10^8\msun}\right]^{\frac{2}{5}}
\left[\frac{\sigma\rmh}{200\kms}\right]^{-2}
\left[\frac{R\lrT}{0.1}\right]^{\frac{2}{5}}
\left[\frac{T}{10\gyr}\right]^{-\frac{2}{5}}. \nonumber \\
\label{eq:calET2calEhAbove}
\end{eqnarray}
Substituting Equation~\eqref{eq:calET2calEhAbove} into
Equation~\eqref{eq:scrFdrainPL}, the corresponding stellar consumption rate due
to the loss-region draining when the MBH mass is above the separatrix can be
approximated in the power-law model as
\begin{eqnarray}
\scrFdrainPL\simeq 
0.735^{\alpha-3}\frac{2\xi(\alpha)}{2\alpha-1} \times 10^{-3}f_T\pyr
\left[\frac{M\bh}{10^8\msun}\right]^{\frac{2\alpha-1}{5}}
\nonumber \\
\left[\frac{\sigma\rmh}{200\kms}\right]^{6-2\alpha}
\left[\frac{R\lrT}{0.1}\right]^{\frac{2\alpha-1}{5}}
\left[\frac{10\gyr}{T}\right]^{\frac{2\alpha-1}{5}}.
\nonumber \\
\label{eq:scrFdrainPLAbove}
\end{eqnarray}

For the convenience of the following analysis, we define
\begin{equation}
\chi(\alpha)= \left\lbrace\begin{split}
1.24^{\alpha-3}\cdot\frac{2\xi(\alpha)}{2\alpha-1},\quad r\rmt\geq 2r\swl, \\
0.735^{\alpha-3}\cdot\frac{2\xi(\alpha)}{2\alpha-1},\quad r\rmt< 2r\swl.
\end{split}\right.
\label{eq:chialpha}
\end{equation}
and
\begin{eqnarray}
\kappa(M\bh,\sigma\rmh) = \left\lbrace\begin{split}
M_{\rm BH,8}^{\frac{2}{3}\alpha-1} \sigma_{\rm h,200}^{6-2\alpha}, 
\quad r\rmt\geq 2r\swl, \\
M_{\rm BH,8}^{\frac{2}{5}\alpha-\frac{1}{5}} \sigma_{\rm h,200}^{6-2\alpha},
\quad r\rmt< 2r\swl.
\end{split} \right.
\label{eq:mbhsigmadrain}
\end{eqnarray}
With these definitions, the stellar consumption rate due to the loss-region draining
in the power-law model can be expressed as
\begin{eqnarray}
\scrFdrainPL & \simeq & 10^{-3}f_T\pyr
\chi(\alpha)\kappa(M\bh,\sigma\rmh) \times \nonumber \\
& & \left[\frac{R\lrT}{0.1}\right]^{\frac{2\alpha -1}{5}}
\left[\frac{10\gyr}{T}\right]^{\frac{2\alpha-1}{5}}. 
\label{eq:scrFdrainPLunited}
\end{eqnarray}
In the following, we fix the consumption time to $T=10\gyr$. 

Similarly as done for the case of the two-body relaxation, we compare the
stellar consumption rates calculated from Equation~\eqref{eq:Fdrain} and those
obtained approximately from the power-law model (Eq.~\ref{eq:scrFdrainPLunited})
as shown in Figure~\ref{fig:f5}. As seen from the figure, the rates evaluated
based on these two approaches are generally consistent with each other which
again verifies the effectiveness of the power-law model in estimating the
stellar consumption rate due to the loss-region draining in nonspherical
potentials.

According to Equation~\eqref{eq:scrFdrainPLunited}, both the correlation between
$\scrFdrain$ and $M\bh$ and that between $\scrFdrain$ and $\alpha$ are
controlled by the terms of $\chi(\alpha)$, $\kappa(M\bh,\sigma\rmh)$, and
$\left(\frac{R\lrT}{0.1}\right)^{\frac{2\alpha -1}{5}}$, and Figure~\ref{fig:f6}
shows the dependence of these three terms on $M\bh$ (left panels) and $\alpha$
(right panels), separately. As done for those correlations shown in
Figure~\ref{fig:f4}, we also conduct linear least square fittings to the data
points shown in each panel of Figure~\ref{fig:f6}. The best fit is indicated by
the red dashed line in each panel and  the best fit form is also labeled there.
The best-fit parameters with their $1\sigma$ uncertainties are also listed in
Table~\ref{tab:t1}.

We use the fitting results to identify the main factors that contribute to the
correlation between $\scrFdrain$ and $M\bh$ and that between $\scrFdrain$ and
$\alpha$ (see Tab.~\ref{tab:t1} and legends in Figs.~\ref{fig:f2} and
\ref{fig:f6}).
\begin{itemize}
\item
As seen from the left panel of Figure~\ref{fig:f2}, the $\scrFdrain$--$M\bh$
correlation has a slope of $0.90$. While according to the left panels of
Figure~\ref{fig:f6}, the correlations between those three inspected terms (in
logarithm) and $\log M\bh$ have slopes of $0.16$, $0.69$, and $0.01$,
respectively. Overall, these three relations combined can well account for the
correlation between $\scrFdrain$ and $M\bh$. Among the three terms,
$\mbhsigmadrain$ dominates the correlation, which scales as $\sim \sigma\rmh^3$
and $\sim \sigma\rmh^3 M\bh^{2/5}$ when $M\bh$ is below and above the
separatrix, respectively, if setting $\alpha=3/2$. In addition, the term
$\chi(\alpha)$ contributes minorly to the correlation. By comparing the data
points in the left panels of Figure~\ref{fig:f6} and the left panel of
Figure~\ref{fig:f2}, it is clearly revealed that the scatter of the correlation
between $\scrFdrain$ and $M\bh$ is also dominated by $\mbhsigmadrain$. 
\item
As seen from the right panel of Figure~\ref{fig:f2}, there exist considerable
scatters among the data points. Given the scatters, the correlation between
$\scrFdrain$ and $\alpha$ appears to be mild. A negative slope ($-0.64$) is
returned when the linear square regression is applied to the data points,
suggesting a decrease of about a factor of $4$ from $\alpha=1$ to $\alpha=2$.
From the right panels of Figure~\ref{fig:f6}, the best-fit slopes of the three
inspected terms (in logarithm) against $\alpha$ are $-0.64$, $0.20$, and
$-0.21$, respectively. Again, these three relations combined together can well
account for that correlation between $\scrFdrain$ and $\alpha$. Among the three
terms, $\chi(\alpha)$ dominates the correlation. The scatter of the correlation
is largely contributed by the term of $\mbhsigmadrain$.
\end{itemize}

\section{Discussions} 
\label{sec:discuss}

\subsection{Overrepresentation of TDEs in E+A/poststarburst galaxies}
\label{sec:discuss:EplusA}

Some recent studies towards TDE host galaxies suggest that TDEs prefer to happen
in those rare E+A/poststarburst galaxies \citep{Arcavi14, French16, French17,
LawSmith17, Graur18, Hammerstein21}. The underlying (physical) explanations for
the preference are still unclear, and there may be alternative characteristics
to distinguish TDE host galaxies from normal ones \citep{LawSmith17}. Here we
discuss whether the preference can be explained by the correlations between the
stellar consumption rate and the inner slope of the galaxy stellar number/mass
density distribution. Since TDE flares can only be observed when the MBH mass is
below the Hills' mass \citep{Hills75}, in which regime the two-body relaxation
mechanism dominates over the stellar orbital precession mechanism (CYL20), we
focus on the correlation between $\scrFlc$ and $\alpha$.

To explore whether the overrepresentation can be explained by the high central
stellar density of the galaxies, \citet{LawSmith17} identified a region in the
Sersic index and MBH mass parameter space which contained ${\sim} 2\%$ of their
reference catalog galaxies but $\geq 60\%$ of those TDE host galaxies. This
means that the averaged TDE rate in those high-Sersic-index galaxies should be
higher than that in their low-Sersic-index counterparts by a factor of ${\sim}
25-48$. From the right panel of Figure~\ref{fig:f1}, the mean value of $\log
\scrFlc$ increases by $1.37$ when $\alpha$ is increased from $1$ to $2$, or
equivalently when $\gamma$ (the inner slope of the galaxy surface brightness
profile) is increased from $0$ to $1$. As also seen from the panel, the
intrinsic scatter of the correlation has no significant dependence on $\alpha$.
Therefore, the averaged consumption rate $\langle\scrFlc\rangle$ is increased by
a factor of ${\sim} 24$ from $\alpha=1$ to $\alpha=2$, which is within a factor
of $2$ of the analysis by \citet{LawSmith17}. Given the limited number of TDEs
in the observational analysis (i.e., 5--10 events), it suggests that the
correlation between $\scrFlc$ and $\alpha$ may be responsible for the overrepresentation
reported by recent studies towards TDE host galaxies.

The overrepresentation may also be contributed by draining of the loss-region
stars in nonspherical potentials, 
as those poststarburst systems may also have a higher degree
of asymmetry in their shapes as compared with generic galaxies. 
The presence of A type stars in the nuclear
regions of those E+A/poststarburst galaxies indicates their relatively young
ages and dynamical states (e.g., \citealt{French17, LawSmith17}), and therefore
a relatively full loss region and a larger draining rate.  
As the loss region draining rate $\scrFdrain$ decreases with increasing time
$T$,
we can define the time $T\rmeq$ at which the loss region draining rate $\scrFdrain
(T=T\rmeq)$ equals to the loss cone refilling rate $f^{\rm tri}\scrFlc$.
We have $\scrFdrain(T)\ga f^{\rm tri}\scrFlc$ at $T\la T\rmeq$
and $\scrFdrain(T)\la f^{\rm tri}\scrFlc$ at $T\ga T\rmeq$. 
By using the scaling dependence of $\scrFdrainPL$ on time $T$ shown in
Equation (\ref{eq:scrFdrainPLunited}) (i.e., $\propto T^{(1-2\alpha)/5}$) 
and the scaling dependence of $\scrFdrain$ and $\scrFlc$ on $M\bh$ shown at
$T=10\gyr$ in
Figures~\ref{fig:f1}--\ref{fig:f2}, we can obtain the time $T\rmeq$ by
\begin{equation}
\left(\frac{2\alpha-1}{5}\right)
\log\left(\frac{T\rmeq}{10\gyr}\right)\simeq 
1.10-\log f^{\rm tri}+1.18\log M_{\rm bh,8}.
\label{eq:Teq}
\end{equation}
For example, we have $T\rmeq\simeq 10\gyr 
({f^{\rm tri}\over 3})^{-0.4} ({M\bh\over3\times 10^7M_\odot})^{2.95}$ if $\alpha=3/2$ in Equation (\ref{eq:Teq}).
Compared with the stellar consumption rate in a system with age $\sim10\gyr$,  the enhancement of stellar consumption rate due to the draining
of the loss-region stars in a young system with dynamical age $T$ can be estimated by a factor of ${\scrFdrain(T)\over \scrFdrain[\min(10\gyr,T\rmeq)]}\simeq [{\min(10\gyr,T\rmeq)\over T}]^{\frac{2\alpha-1}{5}}$. 
If assuming the age $T=0.1\gyr$, and $f^{\rm tri}\simeq 3$ and $\alpha=3/2$
in Equation (\ref{eq:Teq}),
the factor is $\sim 6$ for $M\bh\ga 3\times 10^7M_\odot$ (for which $T\rmeq\ga10\gyr$)
and smaller for lower MBH masses (e.g.\ $\sim 2$ for $M\bh=10^7M_\odot$).
Those estimation obtained from the above analytical scaling
fittings are consistent with those shown in Figure~5 of CYL20. 

Overall, the TDE event rate in the E+A galaxies may be enhanced by a factor of
several tens as compared with generic galaxies. Among them, 
the effect due to larger $\alpha$ or higher central densities of E+A galaxies dominates the rate enhancement, while draining
of the loss-region stars in these dynamically young systems may also contribute
a factor of a few, depending on the dynamical age of the system and the MBH mass.
We expect that the above explaination to the overrepresentation of the TDEs due
to the different contributions will be testable through a significant accumulation of
TDE observations along with observations of their host galaxy properties.

\subsection{Generalization for different types of stars}

\label{sec:discuss:multipop}

In the above analysis, we assume a single stellar population with solar mass
and solar radius.
In reality, however, the stellar system is composed of a spectrum of stars with
different masses $m_\ast$ and radii $R_\ast$. 
For each species in the stellar system, the mass separatrix of MBHs that
can tidally disrupt or directly swallow low-angular-momentum stars
is determined by a comparison of $r_{\rm t}$ and $2r_{\rm swl}$, as
mentioned in Section~\ref{sec:model:drain}.
Different types of stars have different $M\bh$ separatrixes between TDEs and
direct capture events, which depend on
stellar mass and radius by
a scaling factor of $(m_\ast/M_\odot)^{-\frac{1}{2}}
(R_\ast/R_\odot)^\frac{3}{2}$.
In this subsection, we discuss how the mass
spectrum may affect the stellar consumption rate estimations and their correlation tendencies and how
our above scaling analysis can imply for the rates of
different types of stars.

We discuss the effects of the mass spectrum on the stellar consumption rate estimations and their
correlation tendencies through their effects on $\scrFlc$ and $\scrFdrain$ as follows. 
\begin{itemize}
\item $\scrFlc$:
The overall stellar consumption rate due to two-body relaxation, $\scrFlc$, is
not affected significantly by the generalization of the single-mass stellar population assumption (MT99). 
As shown in Appendix A in MT99,
if replacing the single-mass stellar population with an old stellar population
(e.g., the Kroupa initial mass function between $0.08M_\odot$ and $1M_\odot$),
$\scrFlc$ will increase by a factor of $\sim 1.66$, due to the combined
effects of the increased stellar number density and the decreased stellar
diffusion rate.
The consumption rate of a
given type of stars is then roughly proportional to its number fraction among
all the stars, if the mass segregation effect is not significant. However, if the
mass segregation effect is important, then the consumption rate of high-mass
stars may be further enhanced and that of low-mass stars may be weakened. 

For each species in the stellar system, the consumption rate due to two-body
relaxation varies with the changes of the number fraction of the species and
the logarithmic terms $\ln Q$ and $\ln R_0^{-1}(\calE)$ in
Equations~\eqref{eq:EFE}--\eqref{eq:scrFlcPL}.
If we ignore the generally weak variations of the two logarithmic terms, the
$\scrFlc$ rate
correlation tendencies due to two-body relaxation apply effectively to
disrupted or swallowed
stars of different types, including giant stars (with $m_\ast
{\sim} M_\odot$, $R_\ast {\sim} 10$--$1000 R_\odot$) and white dwarfs (with
$m_\ast {\sim} M_\odot$, $R_\ast {\sim} 0.01 R_\odot$). 

\item $\scrFdrain$:
The stellar consumption rate due to loss-region draining, $\scrFdrain$, is
affected by relaxing the single-type stellar population assumption in the
following three aspects. (a) $\scrFdrain$ is
proportional to the number density of stars (corresponding to the $N\rmh$ term
in Eq.~\ref{eq:scrFdrainPL}). 
If the single stellar population with solar
mass and radius considered above in Section~\ref{sec:model:drain} is replaced by
an old population of stars (e.g., the Kroupa initial mass function between
$0.08M_\odot$ and $1M_\odot$), the change of the total stellar number density
will lead to an increase of $\scrFdrain$ by a factor of $5.33$. 
The rate of each given species is 
proportional to the number fraction of the given species.
(b) $\scrFdrain$ can also depend on mass and radius of each type of stars
via their different loss-cone size or the consumption radius $r_{\rm consp}$.
According to Equations (\ref{eq:calET}) and (\ref{eq:scrFdrainPL}), we have
$\scrFdrain\propto {r_{\rm consp}}^{{-2\alpha+6}\over 5}$,
which is $\propto (m_\ast/M_\odot)^\frac{2\alpha-6}{15}$
$(R_\ast/R_\odot)^\frac{-2\alpha+6}{5}$ for MBHs with mass below the mass
separatrix between TDEs and direct capture events (Eq.~\ref{eq:scrFdrainPLBelow}).
For MBHs with mass above the mass separatrix, the variable of $r_{\rm consp}$
in $\scrFdrain$ is $\propto M\bh$ and does not contribute to the
dependence on $m_\ast$ and $R_\ast$ (Eq.~\ref{eq:scrFdrainPLAbove}).
(c) $\scrFdrain$ can depend on the evolutionary ages of the different types
of stars through the factor of $T^\frac{-2\alpha+1}{5}$ in Equation
(\ref{eq:scrFdrainPLunited}).

As for the loss region draining processes, the different stellar species move
in the same triaxial galactic potential and are independent of each other, and
the $\scrFdrain$ rate correlation tendencies with $M\bh$ and $\alpha$ due to
loss-region draining obtained in Section~\ref{sec:model:drain} (see
Eqs.~\ref{eq:scrFdrainPLBelow} and \ref{eq:scrFdrainPLAbove})
can be directly applied to the different stellar species, although the detailed
fit correlation cooefficients could be affected quantitatively by their
differences in the mass separatrix of MBHs between TDEs and direct capture
events.

\end{itemize}

Based on the above analysis, we discuss the implications specifically for the
following types of stars which have different characteristic masses and radii. 
\begin{itemize}
\item Giant stars: For giant stars (with $m_\ast {\sim} M_\odot$, $R_\ast {\sim} 10$--$1000
R_\odot$), the upper mass boundary of MBHs being able to produce TDE
flares increases to much larger values, i.e., $\gg 10^8 M_\odot$.
Based on Equation (\ref{eq:scrFdrainPLBelow}) and including the enlarged
loss-cone size since the transition to giant star from their main-sequence stages, the ratio of $\scrFdrain$ for giant stars to
that for solar-type stars can be estimated by the factor of
$(m_\ast/M_\odot)^\frac{2\alpha-6}{15}
(R_\ast/R_\odot)^\frac{-2\alpha+6}{5}(T/10\gyr)^\frac{-2\alpha+1}{5}f_{\rm g}
\sim 1$, if adopting $\alpha=3/2$, $m_*\sim M_\odot$, the lifetime $T\sim 1\gyr$ for $R_*\sim 10 R_\odot$
giants or $T=0.01\gyr$ for $R_*\sim 1000 R_\odot$ giants, and if the number ratio of
the giants to solar-type stars $f_{\rm g}$ is estimated by $\sim T/10\gyr$. 
Note that the estimates of $\scrFdrain$ for giant stars in the above example are
significantly high, in which the TDE rates for giant stars can be up
to those for solar-type stars.

For TDEs of giant stars, we expect that the correlation between the
TDE rates and the MBH mass is similar as obtained from the single solar-type
stellar population assumption, i.e., dominated by the negative
$\scrFlc$--$M\bh$ correlation at small $M\bh$, and by the positive
$\scrFdrain$--$M\bh$ correlation at large $M\bh$, The transitional MBH mass
between those two correlation tendencies with $M\bh$ should be smaller than
$\sim10^7 M_\odot$ due to the relative increase in loss-region draining rates
$\scrFdrain$.
At the mass range below or above the transitional MBH mass, the analysis of the
correlation tendencies of the TDE rates with $M\bh$ and $\alpha$ are expected
to follow those shown in Section~\ref{sec:model:drain}. 

\item Massive young stars: For massive stars, 
the upper mass boundary of MBHs being able to produce TDE
flares can also increase to much larger values, i.e., $\gg 10^8 M_\odot$.
Note that the increase of $\scrFdrain$ due to the young dynamical age has been
discussed in Section~\ref{sec:discuss:EplusA}.
For massive stars, the dependence of $\scrFdrain$ on
$(m_\ast/M_\odot)^\frac{2\alpha-6}{15} (R_\ast/R_\odot)^\frac{-2\alpha+6}{5}$
is relatively not strong, for example, $\scrFdrain$ can increase by a factor of
$\sim 2$--$4$ (for $m_\ast\sim 10$--$100 M_\odot$ and $R_*\propto
m_\ast^{0.8}$), if adopting $\alpha=3/2$, due to the enlarged loss-cone size. 

The analysis on the correlation tendencies of $\scrFdrain$ and $\scrFlc$ with
$M\bh$ and $\alpha$ and the transitional MBH mass of the correlations for
giant stars above can also be applied to the analysis for TDE samples of
massive young stars (e.g., $T\ll 1\gyr$).

\item Stellar compact remnants (white dwarfs, neutron stars, and stellar-mass
BHs): Stellar compact remnants can generally be swallowed directly by MBHs when
they move sufficiently close to the MBH, with bursts of gravitational waves,
except that white dwarfs can be tidally disrupted by MBHs if $M\bh \lesssim
10^5M_\odot$ (at the lower boundary of or beyond the mass range considered in
this paper). 

The TDE rates of white dwarfs should be dominated by the loss-region refill
rate due to two-body relaxation,
and the rate estimation is subject to the uncertainties in the statistics of
the MBH population at the low-$M\bh$ end and their stellar environments.

For $M\bh> 10^5M_\odot$, if the mass segregation effect of compact objects at
galactic centers is ignored, the direct capture rates of compact objects and
their correlation tendencies follow the similar analysis on $\scrFlc$ and
$\scrFdrain$ shown in Section~\ref{sec:model}, with including the number
fraction of the different types of compact objects. Note that the loss-region
draining rates of the different types of compact objects are irrelevant with
their detailed masses and radii, as their $r_{\rm consp} =2r_{\rm swl}$ are the
same given the same $M\bh$ (see also Eq.~\ref{eq:scrFdrainPLAbove}).

\end{itemize}

\section{Conclusions}
\label{sec:conclusions}

In this work, we study the correlations of the stellar consumption rates by the
central MBHs of galaxies with both the MBH mass $M\bh$ and the inner slope of
the host galaxy stellar number/mass density distribution $\alpha$. The rates of
stellar consumption due to two-body relaxation $\scrFlc$ and stellar orbital
precession in nonspherical potentials $\scrFdrain$ are considered. By exploiting
a simplified power-law model, i.e., considering a single power-law stellar
number/mass density distribution under the Keplerian potential of the central
MBH, we derive approximated expressions for the stellar consumption rates due to
both the mechanisms. Then by inspecting the relative contributions from
different terms in the approximated rate expressions to the correlations, we
identify the dominant factor(s) responsible for both the slopes and scatters of
the correlations. We summarize the main conclusions of this study below.

\begin{itemize}

\item
In both cases of the two-body relaxation in spherical galaxy potentials  and the
loss-region draining in nonspherical potentials, the stellar consumption rates
estimated based on the power-law model $\scrFlcPL$ and $\scrFdrainPL$ are
consistent with the rates estimated based on the full model, i.e., $\scrFlc$ and
$\scrFdrain$, respectively. This not only verifies the effectiveness of using
the simplified power-law model to inspect the correlations, but also provides an
efficient and simple way to estimate the stellar consumption/flaring rates due
to both the mechanisms.

\item
$\scrFlc$ correlates negatively with $M\bh$ while positively with $\alpha$. As
for the $\scrFlc$--$M\bh$ correlation, the best-fit linear relation has a slope
of $-0.28$. Both the slope and the scatter of the correlation are dominated by
the term $\sigma\rmh^3 M\bh^{-1}$ in the approximated expression of the stellar
consumption rate $\scrFlcPL$, where $\sigma\rmh^2\equiv GM\bh/r\rmh$ and the
influential radius of the central MBH $r\rmh$ is defined as the radius within
which the stellar mass equals the MBH mass. As for the $\scrFlc$--$\alpha$
correlation, the best-fit linear relation has a slope of $1.37$. Again, both the
slope and the scatter of the correlation are dominated by $\sigma\rmh^3
M\bh^{-1}$.

\item
$\scrFdrain$ correlates positively with $M\bh$ while negatively with $\alpha$.
The latter correlation appears to be mild due to the large scatter in the
relation between $\scrFdrain$ and $\alpha$. As for the $\scrFdrain$--$M\bh$
correlation, the best-fit linear relation has a slope of $0.90$. Both the slope
and scatter of the correlation are dominated by the term $\mbhsigmadrain$
(Eq.~\ref{eq:mbhsigmadrain}) in the approximated expression of $\scrFdrainPL$,
which scales as $\sim \sigma\rmh^3$ and $\sim \sigma\rmh^3 M\bh^{2/5}$ when the
$M\bh$ is below and above the separatrix, respectively, if set $\alpha=3/2$. As
for the $\scrFdrain$--$\alpha$ correlation, the best-fit linear relation has a
slope of $-0.64$. The term $\chi(\alpha)$ (Eq.~\ref{eq:chialpha}) in
$\scrFdrainPL$ dominates the slope while $\mbhsigmadrain$ dominates the scatter
of the correlation.

\item 
The above correlations of $\scrFlc$ and $\scrFdrain$ serve as the backbones of
the correlation tendencies of the stellar consumption rates at the low-mass
($M\bh\lesssim 10^7\msun$) and the high-mass ($M\bh\gtrsim 10^7\msun$) ranges of
MBHs, respectively.

\item
We use the $\scrFlc$--$\alpha$ correlation to explain the overrepresentation of
TDEs in those rare E+A/poststarburst galaxies found by some recent
observational studies (e.g., \citealt{Arcavi14, French16, French17, LawSmith17,
Graur18, Hammerstein21}). According to the correlation, the expectation value
of $\scrFlc$ is increased by a factor of ${\sim}24$ when $\alpha$ is increased
from $1$ to $2$, or equivalently, when $\gamma$ is increased from $0$
(core-like) to $1$ (cuspy). This factor is broadly consistent with the
overrepresentation factor found by \citet{LawSmith17}, i.e., ${\sim}25$--$48$,
indicating that the preference of TDEs in these rare subclass of galaxies can
be largely explained by the $\scrFlc$--$\alpha$ correlation. Besides the
$\scrFlc$--$\alpha$ correlation, loss-region draining in these dynamically
young systems can also enhance the observed TDE rate by a factor of a few,
depending on the dynamical age and the MBH mass. Future observations of TDEs
and their host galaxy properties are expected to test those different
contribution origins.

\item
The stellar consumption rates and their correlation tendencies are discussed
for different types of stars, including giant stars, massive young stars, and
stellar compact remnants. We find that the estimates of the TDE rates of giant
stars can be high enough to be up to or those of solar-type stars, due
to the large tidal disruption radii of giant stars.  How to distinguish the
different TDEs of giant stars and main-sequence stars in observations deserves
further investigations.

\end{itemize}

With the increasing power of the time domain surveys, the TDE samples are
expanding rapidly in recent years \citep{Gezari21}, which enables the study of
the MBH demographics from a brand new perspective (e.g., \citealt{Ramsden22}).
For example, TDEs illuminate those dormant MBHs, which provides a powerful tool
to study the mass function and occupation fraction of MBHs in quiescent
galaxies, especially at the low-mass end where most, if not all, TDEs occur
\citep{Stone16, Fialkov17}. In addition, the upper mass limit of a MBH that can
tidally disrupt a star depends sensitively on the MBH spin \citep{Beloborodov92,
Kesden12, Leloudas16, Mummery20}. Therefore, the observed MBH mass distribution
near ${\sim}10^8$--$10^9\msun$ from a large sample of TDEs can set strong
constraints on the spin distributions of MBHs inside the galaxy centers.
Moreover, a stellar system containing a binary MBH or a recoiled MBH could
undergo a short period of time during which TDEs promptly happen
\citep{CSMetal11, Stone11, Wegg11,Stone12}. Despite these merits, the results
from TDE observations should be interpreted with cautions when constraining the
MBH demographics, since our study reveals that some bias may be induced by the
different galaxy properties. Therefore, we conclude that a thorough
understanding of the dependence of the TDE rate on both properties of MBHs and
their host galaxies is the prerequisite of the MBH demographic study with TDE
observations.

\section*{Acknowledgments}
This work is partly supported by the National Natural Science Foundation of
China (grant Nos.\ 12173001, 11721303, 11873056, 11690024, 11991052), the
National SKA Program of China (grant No. 2020SKA0120101), National Key Program
for Science and Technology Research and Development (grant Nos.\ 2020YFC2201400,
2016YFA0400703/4), and the Strategic Priority Program of the Chinese Academy of
Sciences (grant No. XDB 23040100).

\end{document}